# Am I Responsible for End-User's Security?

## A Programmer's Perspective

Chamila Wijayarathna
University of New South Wales
Canberra, Australia
c.diwelwattagamage@student.unsw.edu.au

Nalin A. G. Arachchilage
University of New South Wales
Canberra, Australia
nalin.asanka@adfa.edu.au

## ABSTRACT
Previous research has pointed that software applications should not depend on programmers to provide security for end-users as majority of programmers are not experts of computer security. On the other hand, some studies have revealed that security experts believe programmers have a major role to play in ensuring the end-users' security. However, there have been no investigation on what programmers perceive about their responsibility for the end-users' security of applications they develop. In this work, by conducting a qualitative experimental study with 40 software developers, we attempted to understand the programmer's perception on who is responsible for ensuring end-users' security of the applications they develop. Results revealed majority of programmers perceive that they are responsible for the end-users' security of applications they develop. Furthermore, results showed that even though programmers aware of things they need to do to ensure end-users' security, they do not often follow them. We believe these results would change the current view on the role that different stakeholders of the software development process (i.e. researchers, security experts, programmers and Application Programming Interface (API) developers) have to play in order to ensure the security of software applications.

## 1. INTRODUCTION
Computers and Information Technology are integral parts of people's life where most day to day activities of people depend on the use of computers and Information Technology. Not only in people's day to day life, most organizations, ranging from small businesses to governments, use computers and Information Technology to carry out various operations. Most of these involve the use of applications that store or transfer sensitive data of users and organizations, which represents a key target to hackers. Despite the continuous evolution of security technologies, it appears that hackers are still capable of identifying security vulnerabilities of software applications to perform attacks against them. Cyber incidents have been increasing in frequency and cost in recent years, with some resulting in hundreds of millions of dollars in losses [2].

Previous studies have identified that mistakes programmers make while developing applications are a major reason for security vulnerabilities that exist in applications [3, 4]. Programmers come from different expertise/backgrounds and have different levels of experience [7]. Most of them would not be experts of security and would be ignorant of the security of applications they develop [7]. This results in those programmers developing vulnerable applications and therefore Wurster and van Oorschot [10] called programmers as the enemy of security. As a solution to this, some researchers suggest that the security of applications should not rely on programmers who develop those applications [5, 6]. However, currently it is not clear whether or not programmers think that they are responsible for the security of applications they develop. In this study, we are trying to investigate what programmers think on who is responsible for the end-users' security of applications.

We conducted an qualitative experimental study with 40 participants where each participant was asked to implement a secure programming solution. At the end of the experiment, we asked them a couple of questions regarding how their mistakes would result in developing a vulnerable solution. We observed some interesting insights about the programmers' perception on the end-users' security of software solutions they developed.

## 2. BACKGROUND
In the current era of software development, people with different levels of experience and different domains of expertise involve in the software development process as programmers [7]. To meet various expectations of their employers, rather than gaining a more in depth understanding of a specific platform, developers tend to increase the breadth of languages and tools they are able to use [7]. However, security is not a major area that developers try to master as it is considered as a secondary/non-functional requirement in the software development process [10]. Therefore, most programmers who are involved in the software development process are not security experts [7, 10]. Furthermore, programmers believe that programmes they develop are not security critical even when those are [10]. This results in programmers developing applications that contain security vulnerabilities.

There are many examples in literature that investigate security vulnerabilities, which got introduced into applications due to mistakes made by programmers [3, 4]. Georgiev et al. [4] identified that many security critical applications such as Amazon EC2 Java library, Amazon's and Paypal's merchant SDKs, osCommerce, ZenCart and Uber-Cart contained security vulnerabilities due to mistakes made by programmers while developing those applications. In a study that inspected 13500 android applications, Fahl et al. [3] revealed that a considerable amount of those apps are vulnerable to Man In The Middle Attacks. According to authors, the reason for these vulnerabilities is the mistakes made by programmers while using SSL/TLS libraries and APIs.

Due to this, there is an attitude among researchers that programmers are the weakest link [6, 10]. Therefore, some have suggested that the responsibility of implementing security should not be given to the programmer. Green and Smith [6] suggest that implementing cryptography related code should not be given to programmers who are not security experts. They suggest that security related functionalities should be implemented and embedded into Application Programming Interfaces (APIs), and allow developers to utilize those functionalities through those APIs, so that non-expert programmers do not get to touch security related code. Furthermore, Gorski and Iacono [5] suggest that the security of end-users of an application should not depend on programmers who are not security experts. They suggest that security experts should develop security APIs and provide interfaces for programmers to use those functionalities without implementing those functionalities on their own. Moreover, Wijayarathna et al. [9] suggest that security APIs that provide security functionalities should be designed in a way such that the security of applications that are developed using those APIs should not depend on programmers who use those security APIs to develop applications.

However, it seems that security experts have a different opinion to this on how programmers should be involved in the development process. In most software development organizations, there is a seperate group of security experts who overlook the security aspects of applications that the organization develops [8]. Thomas et al. [8] revealed that security experts expect programmers to involve and contribute in security development process even though they agree that the involvement of non-security expert programmers can result in developing vulnerable applications.

Even though both researchers and security experts have expressed their opinion on whether programmers should be given the responsibility of the end-users' security, there has been no investigation on what is the programmer's perception on this. Therefore, in this work, we are trying to understand the programmers' perception on how end-users' security would be affected by the way they implement a secure application.

## 3. METHODOLOGY

The study was designed to investigate what is programmer's perception about end-users' security of the applications they develop. This study was approved by the Human Research Ethic Committee of our university.

On a high level, we recruited programmers and asked them to complete a simple task that involves implementing a secure programming solution. Having completed a task before answering questions gives them a particular context to refer to while answering questions. Furthermore, it helps to minimize the limitations that arise due to participants' memorability and recalling capacity of their usual actions when they answer surveys. Once they completed the task, we asked a couple of questions from them regarding the security of the programme they developed and we asked them how the security of the programme would depend on the way they implemented the task.

We conducted the experiment remotely as it was not feasible for us to get programmers to come to a lab to do the study. Therefore, we recruited programmers with Java programming experience from GitHub to participate in this study [1]. We extracted publicly available email addresses of Java developers with significant contributions to Java projects and sent emails inviting them to participate in our study. We offered them with a $15 Amazon gift voucher as a token of appreciation for the participation. In the invitation email, we included a link to sign up for the study. Furthermore, we informed them that participation is voluntary and participants can withdraw from the study at any time. Once people signed up, we filtered out those who did not have any software development experience since our target sample for the study was software developers. Sign up form required participants to enter their name and email address, which were required to send the study material to them. However, such personally identifiable information of the participants were removed from the final data set which we used for the analysis. A total of 40 programmers completed the study successfully.

We used 4 programming tasks where each task required programmers to implement a secure programming solution. Each programming task was completed by 10 participants. We used 4 tasks rather than using a single task to avoid results being biased to a particular context of security. For each task, we asked programmers to use a specific security API, so we can get insights about whether or not programmers delegate the responsibility of end-user's security to the security API. Following are summaries of 4 tasks we used.

- Task 1 : Embedding authentication to a Java Servelet web application using Google Authentication API.

- Task 2 : Securing the password storage of a Java Servelet web application by hashing passwords with SCrypt hashing algorithm. We specifically asked them to use Bouncycastle API to achieve this.

- Task 3 : Fix Cross Site Scripting (XSS) vulnerabilities of a web application using OWASP ESAPI.

- Task 4 : Integrate Transport Layer Security (TLS) into a simple Java socket communication. We specifically asked them to use Java Secure Socket Extension(JSSE) API to achieve this.

Once each participant signed up by completing the sign up form and consented to participate in the study, we sent them details of the programming task to do and code artefacts to use. Participants completed the task remotely on their own computers and we suggested them to complete the task in a time comfortable to them. Once a participant completed the task, they were asked to send their source code so that we can verify whether they have actually spent time trying to complete the task. Then they had to answer following two questions, which we shared through Google forms, based on their experience.

- Q1 : Do you think the security of the end user of the application you developed depends on how you completed the task? Or does it depend only on the security API you used?
  - The security of the programme solely depends on the way I implemented it
  - The security of the programme depends on the way I implemented it as well as on the security API
  - The security of the programme solely depends on the API used
- Q2 : If you think security of the end user depends on how you completed the task, in which ways does it depend?

Results we collected for the 2nd question were qualitative and therefore, those results were coded by two coders independently using NVivo[1]. When coding was completed, coders compared each individual code, and discussed and resolved disagreements.

## 4. STUDY RESULTS

Table 1 shows the results we obtained from question 1. 75% of the participants (n=30) believed that they are at least partially responsible for the security of the programme they developed. This result shows that majority of programmers represented by our sample (more than 50%) think that the security of programmes they develop depends on the way they implement the programme, not only on the security APIs and security tools they use ($p<0.005$).

**Table 1: Results of Q1**

| Response | Number of Participants |
|---|---|
| The security of the programme was solely depends on the way I implemented it | 6 |
| The security of the programme depends on the way I implemented it as well as on the security API | 24 |
| The security of the programme was solely depends on the API used | 10 |

Qualitative analysis revealed 12 codes related to developers' perception on their responsibility on the security of programmes they developed and how their decisions would result in better/worse security. We categorized these codes into 3 themes and hereafter we discuss the identified codes under those themes.

[1]https://www.qsrinternational.com/nvivo/

### 4.1 Who is responsible for security?

Even though previous research has suggested that applications should not depend on programmers who are not security experts [5], participants of our study expressed that security of the applications they developed was reliant on themselves. It was apparent in the results that programmers are aware that they are responsible for the security of applications they develop, which is contradictory to the claim by previous researchers that programmers are ignorant of security [7].

Majority of the participants described how the way they developed the application would affect security of it. They mentioned that mistakes they made while developing the application would result in security breaches. For example, participants who completed the task to fix an XSS vulnerability mentioned that if they missed a location that should be fixed, it will result in the application being vulnerable for XSS attacks. They mentioned that they need to use security APIs correctly in order to ensure security of their applications.

However, majority of programmers believed (n=34) that security APIs they use while developing an application should also take some responsibility in making sure the applications are secure. They believed that while APIs are being implemented correctly, they should also take responsibility in minimizing the mistakes that programmers make that can result in security vulnerabilities. They mentioned that a good level of abstraction of APIs and explanatory API documentation would contribute to enhance the security of applications.

### 4.2 What should programmers do to ensure security?

Participants stated what programmers need to do to ensure security of applications they develop. Participants suggested that they need to follow standards and techniques for developing security applications. For example, one participant mentioned that programmers need to follow standards of handling passwords (eg : use byte arrays instead of Strings to store passwords ) to ensure security.

Furthermore, participants mentioned the importance of testing the application in order to ensure security. However, it was apparent that most participants have not done that in the experiment due to their lack of knowledge and since it takes considerable amount of time. This made them to be less confident about the security of the application they developed.

Participants also mentioned that programmers have to apply new things on their own in addition to functionalities provided by APIs in order to ensure security of applications. Participants who mentioned this believed that functionalities provided by security APIs alone are not sufficient to ensure security. They highlighted the importance of going the extra mile as programmers.

An interesting observation was that participants mentioned these things they should have followed, because they did not follow them, even though they knew the importance of following these while developing an application. Lack of competency of programmers was one of the main reasons for not following standards and techniques of secure development. Extra time required for testing the applications was a main reason for programmers to not test the programme.

## 4.3 What programmers cannot control?

Despite of the above mentioned facts, participants mentioned that correctly implementing security is not easy. Participants suggested that even though APIs implement most of the low level details of the functionalities and gives a high level view for programmers, using those can be still difficult for programmers. Therefore, this supports the claim that previous researchers have stated - applications should not rely on programmers to implement security [5].

One participant suggested that security should be implemented by experienced programmers. He elaborated saying that *"A very good API is no good in the hands of an inexperienced developer. In the end, it's still to the developer to use the tools provided by the API in the way they were meant to be used, and adapt this way to his use case"*.

## 5. DISCUSSION

We identify 3 main interesting findings of this study which we have summarized below.

- Majority of programmers are not ignorant of security. They know that security depends on them.
- Programmers have an idea on what they need to do to ensure security.
- Despite above, programmers find it difficult to ensure security of applications they develop.

Previous research has taken the direction of not depending on programmers to ensure end-user security [5, 9]. Previous research argues that security APIs should be designed to ensure security on their own without depending on programmers [5, 9]. However, our results suggest that the correct way to go ahead would be to involve programmers in the process by informing them. Previous research has stressed that educating and training programmers on security is not a scalable solution for the ever evolving diverse body of programmers [10]. We suggest that security APIs and other programming tools should guide and inform programmers into doing the right thing. Since programmers seems to know what they need to do, a little help from tools and security APIs would help them to practice it and would result in more secure applications. Furthermore, this results suggest that security experts of organizations should get programmers involved for ensuring end-user security of applications, rather than taking the burden on their own. However, we still believe that security API/tool developers and security experts have a major role to play in this.

Due to the recruitment methodology we used (recruitment via GitHub with $15 reward), results were affected from the self-selection bias. Therefore, our results represent developers who are motivated enough to participate in a research study and spend their time. Acar et al. [1] previously identified that programmers volunteer for similar experiments are more active in GitHub compared to other developers. However, we used this method since it allows us to get a diverse and geographically distributed sample of programmers compared to other available methods. Nevertheless, because of this limitations, these results should be interpreted in this context.

## 6. CONCLUSION

In this study, we conducted a qualitative experimental study with 40 software developers to investigate what is the programmers' perception on their responsibility for the security of applications they develop. In the experiment, participants completed a task where they had to implement a secure programming solution and then they answered couple of questions based on their experience. Through the data we collected, we were able to identify some interesting insights on what programmers think about their responsibility on the end-users' security of applications they develop.

We believe these findings will contribute to better understand how programmers perceive end-user security of applications they develop and would help researchers, security API and tool developers, and security experts in supporting programmers to minimize security errors they make.